\documentclass[referee]{mn2e}
\usepackage{graphicx}
\usepackage{epsfig}
\usepackage[colorlinks,linkcolor=red]{hyperref}
\usepackage{color}
\usepackage{txfonts}

\begin{document}
\title[Evolving NS+He star systems to IMBPs]
{Evolving neutron star+helium star systems to intermediate-mass binary pulsars}
\author[Tang, Liu and Wang.]
{W. Tang$^{\rm 1,2,3,4}$\thanks{E-mail:tangwenshi@ynao.ac.cn}, D. Liu$^{\rm 1,2,3,4}$\thanks{E-mail:liudongdong@ynao.ac.cn}, B. Wang$^{\rm 1,2,3,4}$\thanks{E-mail:wangbo@ynao.ac.cn}\\
$^1$Yunnan Observatories, Chinese Academy of Sciences, Kunming 650216, China\\
$^2$Key Laboratory for the Structure and Evolution of Celestial Objects, Yunnan Observatories, Chinese Academy of Sciences,\\
Kunming 650216, China\\
$^3$University of Chinese Academy of Sciences, Beijing 100049, China\\
$^4$Center for Astronomical Mega-Science, Chinese Academy of Sciences, Beijing 100012, China}

\date{}
\pagerange{\pageref{firstpage}--\pageref{lastpage}} \pubyear{2019}
\maketitle

\label{firstpage}

\begin{abstract}\label{0. abstract}
Intermediate-mass binary pulsars (IMBPs) are composed of neutron stars (NSs) and CO/ONe white dwarfs (WDs). It is generally suggested that IMBPs evolve from  intermediate-mass X-ray binaries (IMXBs). However, this scenario is difficult to explain the formation of IMBPs  with orbital periods ($P_{\rm orb}$) less than 3\,\rm d. It has recently been proposed that a system consisting of a neutron star (NS) and a helium (He) star can form IMBPs with $P_{\rm orb}$ less than 3\,\rm d (known as the NS+He star scenario), but previous works can only cover a few observed sources with short orbital periods. We aim to investigate the NS+He star scenario by adopting different descriptions of the Eddington accretion rate ($\dot M_{\rm Edd}$) for NSs and different NS masses ($M_{\rm NS}$) varying from $1.10\,\rm M_{\odot}$ to $1.80\,\rm M_{\odot}$. Our results can cover most of the observed IMBPs with short orbital periods and almost half of the observed IMBPs with long orbital periods. We found that $\dot{M}_{\rm Edd}$$\propto$$M_{\rm NS}$$^{\rm -1/3}$ could match the observations better than a specific value for all NSs. We also found that the final spin periods of NSs slightly decrease with the initial $M_{\rm NS}$. The observed parameters of PSR J0621+1002, which is one of the well-observed IMBPs whose pulsar mass has been precisely measured, can be reproduced by the present work.
\end{abstract}


\section{Introduction} \label{1. Introduction}
Pulsars are fast-spinning neutron stars (NSs) with spin periods varying from milliseconds to several seconds, which are observed not only as single stars but also as binary systems (e.g. Wang et al. 2011). Pulsars are of great importance in many scientific studies, such as gravitational physics, stellar physics, binary evolution, etc (e.g. Kramer 2009). The binary pulsars with helium (He) WD companions ($0.15\,\rm M_{\odot}$ $<$ $M_{\rm WD}$ $<$ 0.45$\,\rm M_{\odot}$) and short spin periods ($P_{\rm s}$ $<$ 30\,\rm ms) are named as low-mass binary pulsars (LMBPs) (e.g. Camilo et al. 1996; Tauris et al. 1999), whereas the binary pulsars with CO/ONe WD companions ($M_{\rm WD}$ $> 0.45\,\rm M_{\odot}$) and relatively long spin periods (15$-$200\,\rm ms) are named as intermediate-mass binary pulsars (IMBPs) (e.g. Camilo et al. 1996, 2001). 

The LMBPs have been suggested to originate from low-mass X-ray binaries (LMXBs; e.g. Alpar et al. 1982; Radhakrishnan \& Srinivasan 1982; Bhattacharya \& van den Heuvel 1991). In this case, low mass companion stars transfer H-rich material to NSs when companions fill their Roche lobe. During this process, NSs can accrete enough material from low mass companions to be spun up to millisecond periods. However, the IMBPs have quite different properties compared with LMBPs, for example, IMBPs have more massive companions, longer spin periods (up to a few hundred milliseconds), higher period derivatives and higher surface magnetic fields compared with LMBPs (e.g. Li 2002). This implies that the formation channel of IMBPs may be different from that of LMBPs. It is generally suggested that most of IMBPs evolve from IMXBs which consist of NSs and intermediate-mass companions (2.0$–$10.0$\,\rm M_{\odot}$) (e.g. van den Heuvel 1975; Liu et al. 2018). Tauris et al. (2000) studied the IMXB scenario by using the “isotropic re-emission mode''. However, their results cannot account for those IMBPs with orbital periods ($P_{\rm orb}$) less than 3\,\rm d. In order to explain those IMBPs with short orbital periods, some alternative formation scenarios of IMBPs have been put forward, e.g. the ONe WD+He star scenario (see Liu et al. 2018) and the NS+He star scenario (see Chen \& Liu 2013).

In the ONe WD+He star scenario, the ONe WD will experience an accretion-induced collapse (AIC) process after accreting He-rich matter and growing in mass close to the Chandrasekhar mass limit (Liu et al. 2018; Wang 2018). This leads to the formation of a NS accompanied by an evolved He star. Then the He star may fill its Roche-lobe again and transfer He-rich matter onto the NS, leading to a recycling process for the NS. Finally, an IMBP will be formed when the He star evolves into a CO/ONe WD (see Liu et al. 2018). In the NS+He star scenario, the initial companion star is a He main sequence star. The He star fills its Roche-lobe when its central He is exhausted and begins to transfer He-rich matter to the NS, leading to a recycling process for the NS. Eventually, the system may evolve into an IMBP with short orbital period (see Chen \& Liu 2013). However, Chen \& Liu (2013) only considered the case of $M_{\rm NS} = 1.4\,\rm M_{\odot}$ and simply set the $\dot M_{\rm Edd} = 3 \times 10^{\rm -8}\,\rm M_{\odot}\,\rm yr^{\rm -1}$ for all  NSs. In this case, only a few observed IMBPs with short orbital periods could be reproduced. In this work, we aim to investigate the NS+He star scenario for the formation of the observed short orbital period IMBPs by considering different descriptions of $\dot M_{\rm Edd}$ for NSs and different $M_{\rm NS}$. 

 This paper is organized as follows. In Sect.\,2, we present the methods for the binary evolution of NS+He star systems. In Sect.\,3, we will demonstrate our results. We will give some discussions in Sect.\,4 and a short summary in Sect.\,5. 
 
\section{numerical code and methods}\label{2 Numerical Code and Methods}  

During the binary evolution, the donor star may fill its Roche lobe, and then transfer He-rich matter onto the surface of the NS. In our calculations, we employ the prescription provided by Tauris et al. (2013) to investigate the mass growth rate of the NS ($\dot{M}_{\rm NS}$), written as:
\begin{equation}
\dot{M}_{\rm NS}=(|\dot{M}_{\rm 2}|-\max[|\dot{M}_{\rm 2}|-\dot{M}_{\rm Edd},0])\cdot e_{\rm acc}\cdot k_{\rm def},
\end{equation}
in which $\dot{M}_{\rm 2}$ is the mass transfer rate of the He star, $e_{\rm acc}$ is the fraction of the transferred matter from the He star that is actually accumulated to the NS, and $k_{\rm def}$ is the ratio of gravitational mass to the rest mass of the accreted material. The parameter $e_{\rm acc}\cdot k_{\rm def}$ is taken as a free parameter (i.e. accretion efficiency), which is used to describe the mass-growth process of the NS. We set $e_{\rm acc}\cdot k_{\rm def}=1$ for a He star donor (e.g. Tauris et al. 2011; Lazarus et al. 2014). Generally, $\dot{M}_{\rm 2}$ is larger than $\dot{M}_{\rm Edd}$. Thus the different methods for calculating $\dot{M}_{\rm Edd}$ may have significant influence on our results. For He accretion,  $\dot{M}_{\rm Edd}$ is set to be $3\times 10^{\rm -8}\,\rm M_{\odot}\,\rm yr^{\rm -1} $ (case 1; e.g. Dewi et al. 2002; Chen et al. 2011) or alternatively relates to the NS mass (case 2; e.g. Tauris et al. 2013; Liu et al. 2018):
\begin{equation}
\dot M_{\rm Edd}=4.6\times 10^{\rm -8}\cdot M^{\rm -1/3}_{\rm NS} \,\rm M_{\odot} yr^{\rm -1}.
\end{equation}
We will compare the influence of these two cases in our calculations. 

The accreted material onto the NS ( $\Delta M_{\rm NS}$) may recycle the NS. We assume that the initial spin angular momentum of the NS is negligible (e.g. Liu et al. 2018). During the recycling process, Chen \& Liu (2013) adopted the prescription provided by Liu \& Chen (2011) to calculate the spin period of the NS, which could be written as:
\begin{equation}
P_{\rm s}= \frac{\rm 2\pi \,I}{\Delta M_{\rm NS} \cdot \sqrt{M_{\rm NS}\cdot R \cdot \rm G}},
\end{equation}
where $P_{\rm s}$ is in unit of s, $\rm I \approx 10^{\rm 45}$g cm$^{\rm 2}$ is the momentum of inertia of the NS. $\rm G$ is the gravitational constant and $R$ is the radius of the NS. Besides, Tauris, Langer \& Kramer (2012) also obtained a simple convenient expression written as:
\begin{equation}
P_{\rm s}\approx 0.34\times(\Delta M_{\rm NS}/\rm M_{\odot})^{\rm -3/4},
\end{equation}
in which $P_{\rm s}$ and $\Delta M_{\rm NS}$ are in units of ms and $\rm M_{\odot}$, respectively. We also study the influence of these two prescriptions on the final results. 

 We evolved a large number of NS+He star binaries for the formation of IMBPs. In our calculations, the initial masses of He stars ($M^{\rm i}_{\rm He}$) and NSs ($M^{\rm i}_{\rm NS}$) range from $0.67-2.20\,\rm M_{\odot}$ and $1.10-1.80\,\rm M_{\odot}$, respectively; the initial orbital periods ($\log$ $P^{\rm i}_{\rm orb}/\,\rm d$) range from  $-$1.3 to 3.3. Our initial parameters are determined based as follows: (1) We assume that the He stars will evolve into WDs when the initial masses of He stars are less than $2.2\,\rm M_{\odot}$, otherwise they will collapse into NSs or black holes (Avila Reese 1993). (2) Kiziltan et al. (2013) analyzed 18 observed NS+WD binary systems via the Bayesian approach, and found that 95\% of NS masses are in the range of $1.13-2.07\,\rm M_{\odot}$ after accreting  matter of maximum $0.1-0.2\,\rm M_{\odot}$. Thus we set the initial NS masses ranging from $1.10-1.80\,\rm M_{\odot}$.
 
\section{Results} \label{3. Results}
 
\subsection{A typical example for binary evolution}
\begin{figure*}
\epsfig{file=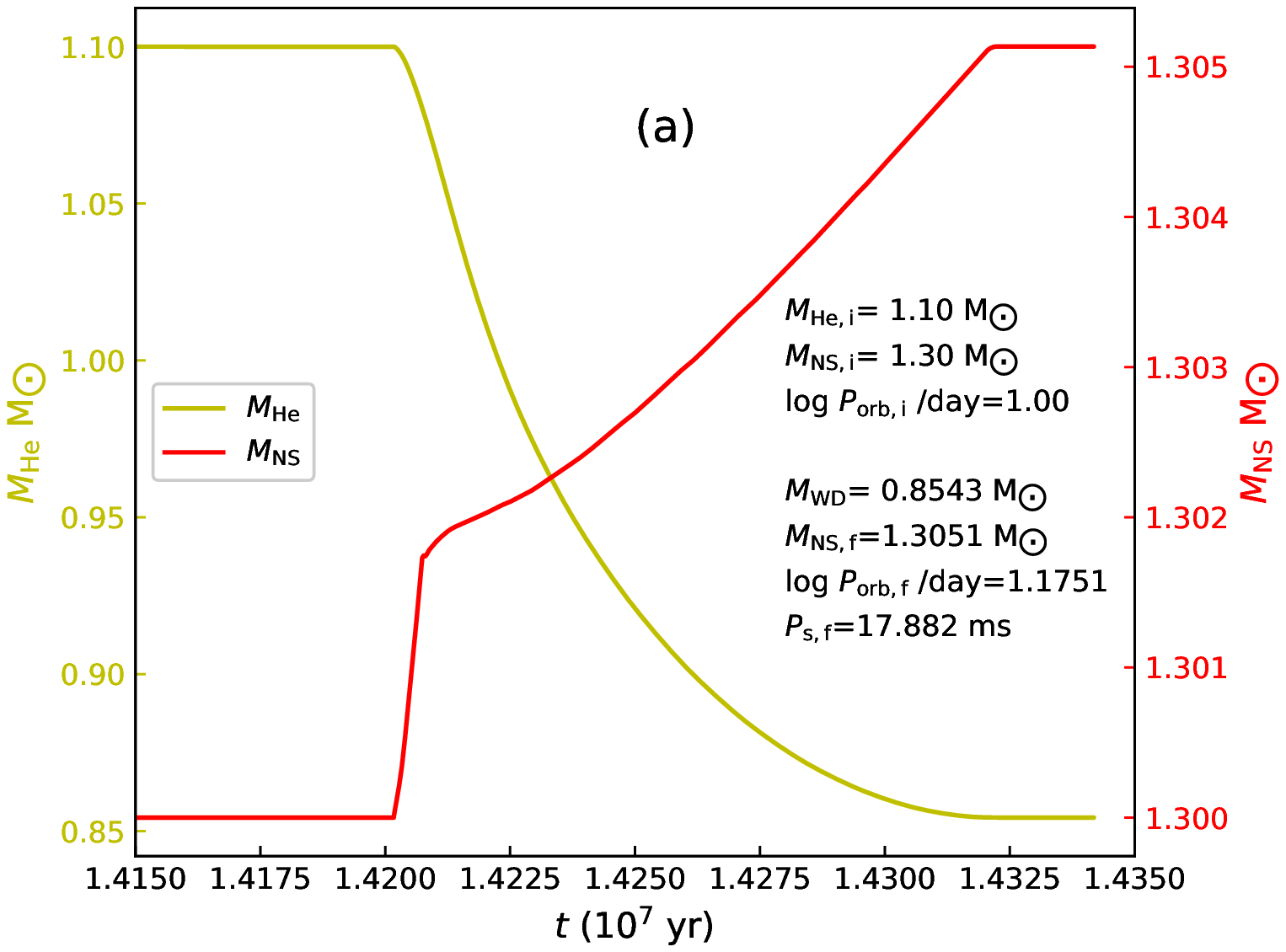,angle=0,width=9cm}\ \
\epsfig{file=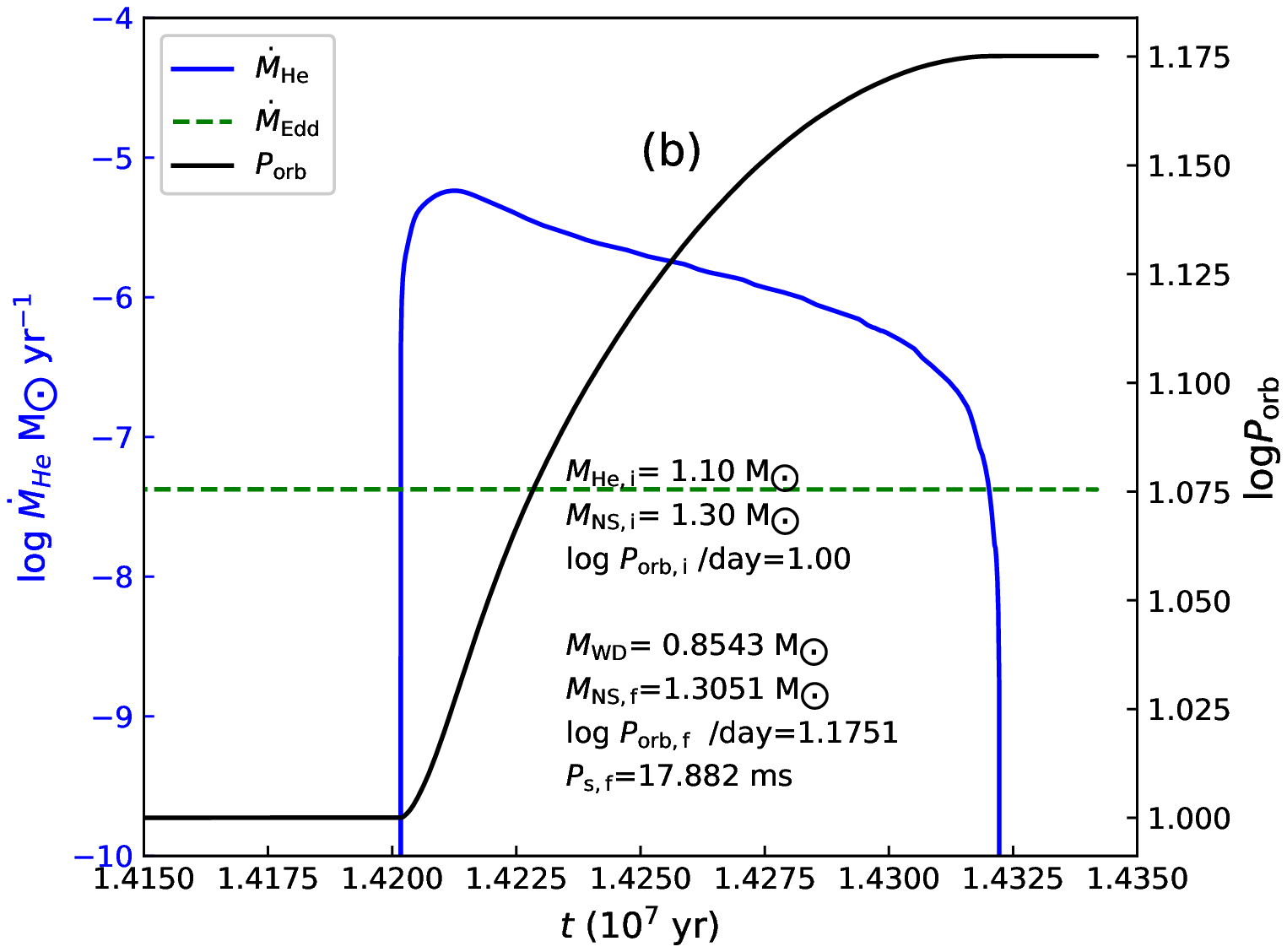,angle=0,width=9cm}\ \
\epsfig{file=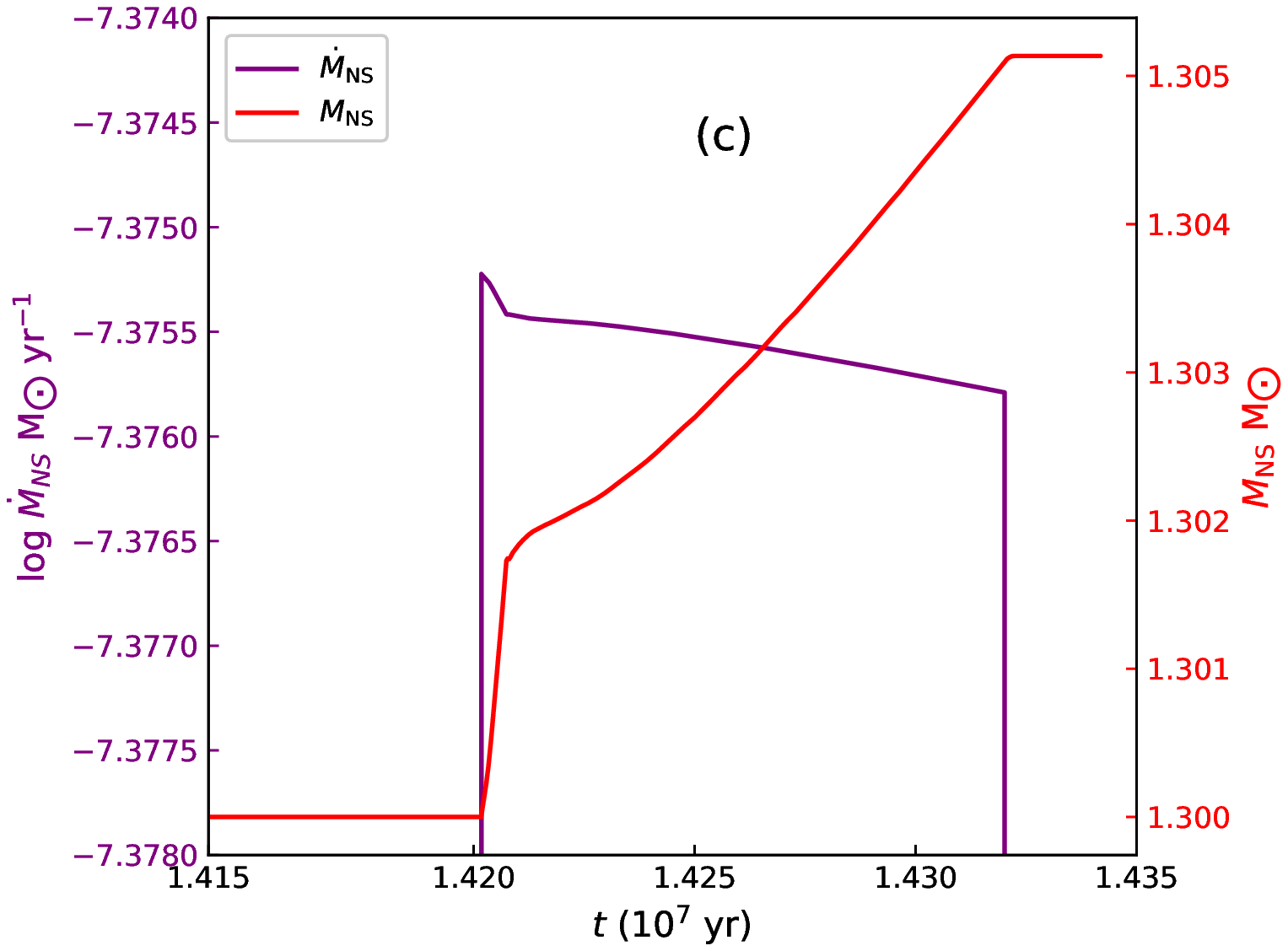,angle=0,width=9cm}\ \
  \caption{A typical example of the evolution of a NS+He star system until the formation of an IMBP. Panel\,\rm (a) presents the evolution of $M_{\rm He}$ (the yellow curve) and $M_{\rm NS}$ (the red curve) as a function of time. Panel\,\rm (b) shows evolution of  $\dot M_{\rm He}$ (the blue curve) and $P_{\rm orb}$ (the black curve). The $\dot M_{\rm Edd}$ is also shown by  green curve. Panel\,\rm (c) displays the enlarged details of the accretion rate of the NS (the purple curve) and the evolution of $M_{\rm NS}$ (red curve).}
\end{figure*}

 In Fig.\,\rm 1, we present the detailed evolutionary sequences of a NS + He star binary system with $M^{\rm i}_{\rm He}= 1.10\,\rm M_{\odot}$,  $M^{\rm i}_{\rm NS}= 1.30\,\rm M_{\odot}$ and $P^{\rm i}_{\rm orb}= 10\,\rm d$. The $\dot M_{\rm Edd}$ is calculated as described in Eq.\,\rm (2). After about 14.2\,\rm Myr, the He star begins to fill its Roche lobe and transfer He-rich matter onto the NS, leading to a recycling process for the NS. $\dot{M}_{\rm 2}$ quickly increases to be significantly higher than $\dot{M}_{\rm Edd}$. In this case, the majority of the transferred material is blown away from the binary at a rate of ($\dot{M}_{\rm 2}-\dot{M}_{\rm Edd}$) driven by the radiation pressure of the NS, and the NS grows in mass at a rate of $\dot{M}_{\rm Edd}$. About $1.25\times 10^{\rm 5}\,\rm yr$ later, the He star exhausts its He-shell and evolves to a CO WD, then the mass transfer process terminates. Eventually, the binary evolves to an IMBP consisting of a $1.3051\,\rm M_{\odot}$  NS with a spin period of 17.82\,\rm ms and a $0.8543\,\rm M_{\odot}$ CO WD. The orbital period expands to be about 14.9658 d. In addition, there is a sudden change of the NS mass growth at around 14.21\,\rm Myr in Fig.\,1\,\rm (a),  which is caused by the sudden decrease of the NS accretion rate presented in  Fig.\,1\,\rm (c).
 \subsection{Parameter space}
 \begin{figure}
\begin{center}
\epsfig{file=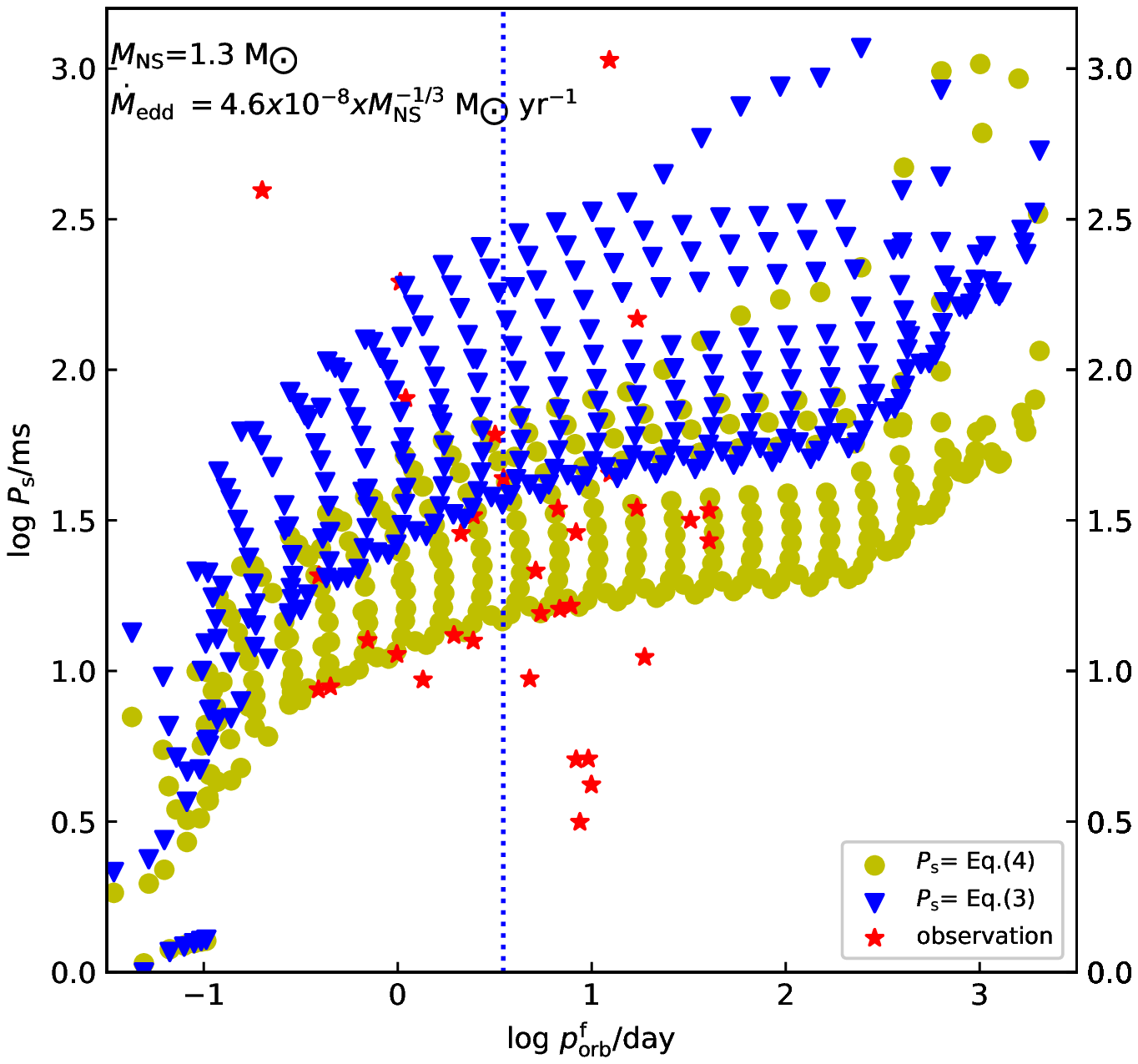,angle=0,width=10.5cm}
 \caption{Display the parameter space of IMBPs in the $\log P^{\rm f}_{\rm orb} –\log P_{\rm s}$ panel, in which initial NS masses are set to be $1.30\,\rm M_{\odot}$ and $\dot M_{\rm Edd}$ is calculated based on Eq.\,\rm (2). The red stars are the observed IMBPs; the blue triangles represent $P_{\rm s}$ according to Eq.\,\rm (3) and the yellow circles represent $P_{\rm s}$ according to Eq.\,\rm (4). The dotted vertical line stands for the $P_{\rm orb} \approx 3\,\rm d$.}
  \end{center}
\end{figure}

Fig. 2 displays the parameter space of IMBPs  in the $\log P^{\rm f}_{\rm orb} –\log P_{\rm s}$ plane. In this case, NS masses are set to be $1.30\,\rm M_{\odot}$ and $\dot M_{\rm Edd}$ is calculated according to Eq.\,\rm (2). The difference is that we utilize different methods to calculate $P_{\rm s}$, i.e. we adopt Eq.\,\rm (3) and Eq.\,\rm (4) to calculate $P_{\rm s}$. The observed IMBPs listed in Table 1 are plotted by red stars. From this figure, we can see that the spin periods of NSs calculated by Eq.\,\rm (4) are shorter than those calculated by Eq.\,\rm (3). This originates from different accretion radii used in these two methods. The magnetospheric boundary is employed in Eq.\,\rm (4) (see Eq. (12) in Tauris, Langer \& Kramer 2012), whereas the real radius of a NS is used in Eq.\,\rm (3). In the following parts of our work, we will use Eq.\,\rm (4) to calculate $P_{\rm s}$. We will discuss about these two methods of calculating $P_{\rm s}$ in Sect.\,4.

\begin{table*}
\begin{center}
 \caption{The parameters of the 34 observed IMBPs taken from the ATNF Pulsar Catalogue (see Manchester et al. 2005; \url {http://www.atnf.csiro.au/research/pulsar/psrcat})}
   \begin{tabular}{ccccccccc}
\hline \hline
 $\rm No.$ & $\rm Pulsars$ & $\rm P_{\rm spin}/ms$ & $\rm P_{\rm orb}/d$\\
\hline
$1$ & $\rm J0621+1002$ & $28.85$ & $8.32$\\
$2$ & $\rm B0655+64$ & $195.7$ & $1.03$\\
$3$ & $\rm J0721-2038$ & $15.54$ & $5.46$\\
$4$ & $\rm J0900-3144$ & $11.11$ & $18.74$\\
$5$ & $\rm J1022+1001$ & $16.45$ & $7.81$\\
$6$ & $\rm J1101-6424$ & $5.11$ & $9.61$\\
$7$ & $\rm J1141-6545$ & $393.90$ & $0.20$\\
$8$ & $\rm J1157-5112$ & $43.59$ & $3.51$\\
$9$ & $\rm J1227-6208$ & $34.53$ & $6.72$\\
$10$ & $\rm J1244-6359$ & $147.27$ & $17.17$\\
$11$ & $\rm J1337-6423$ & $9.42$ & $4.79$\\
$12$ & $\rm J1420-5625$ & $34.12$ & $40.29$\\
$13$ & $\rm J1435-6100$ & $9.35$ & $1.35$\\
$14$ & $\rm J1439-5501$ & $28.63$ & $2.12$\\
$15$ & $\rm J1454-5846$ & $45.25$ & $12.42$\\
$16$ & $\rm J1525-5545$ & $11.36$ & $0.99$\\ 
$17$ & $\rm J1528-3146$ & $60.82$ & $3.18$\\
$18$ & $\rm J1614-2230$ & $3.15$ & $8.69$\\
$19$ & $\rm J1727-2946$ & $27.08$ & $40.31$\\
$20$ & $\rm J1748-2446J$ & $80.34$ & $1.10$\\
$21$ & $\rm J1748-2446N$ & $8.67$ & $0.39$\\
$22$ & $\rm J1750-2536$ & $34.75$ & $17.14$\\
$23$ & $\rm J1757-5322$ & $8.87$ & $0.45$\\
$24$ & $\rm J1802-2124$ & $12.65$ & $0.70$\\
$25$ & $\rm J1807-2459B$ & $4.19$ & $9.96$\\
$26$ & $\rm J1933+1726$ & $21.51$ & $5.15$\\
$27$ & $\rm J1943+2210$ & $5.08$ & $8.31$\\
$28$ & $\rm J1949+3106$ & $13.14$ & $1.95$\\
$29$ & $\rm J1952+2630$ & $20.73$ & $0.392$\\
$30$ & $\rm J2045+3633$ & $31.68$ & $32.30$\\
$31$ & $\rm J2053+4650$ & $12.59$ & $2.45$\\
$32$ & $\rm J2145-0750$ & $16.05$ & $6.84$\\
$33$ & $\rm J2222-0137$ & $32.82$ & $2.45$\\
$34$ & $\rm B2303+46$ & $1066.37$ & $12.34$\\

\hline \label{1}
\end{tabular}
\end{center}
\end{table*}

\begin{table*}
\begin{center}
 \caption{The comparison between observation and model results of PSR J0621+1002:}
   \begin{tabular}{ccccccccc}
\hline \hline
 $ $ & $\rm Observation$ & $\rm Model\,\ results$\\
\hline
$\rm M_{\rm NS} \rm / M_{\odot}$ & $\rm 1.70^{\rm +0.32}_{\rm -0.29}$ & $1.7000075$\\
$\rm M_{\rm WD} \rm/ M_{\odot}$ & $\rm 0.97^{\rm +0.27}_{\rm -0.15}$ & $0.95$\\
$\rm P_{\rm orb} \rm{/ day}$ & $ 8.32$ & $8.13$\\
$\rm P_{\rm s}\rm / ms$ & $28.85$ & $28.55$\\

\hline \label{2}
\end{tabular}
\end{center}
\end{table*}

\begin{figure}
\begin{center}
\epsfig{file=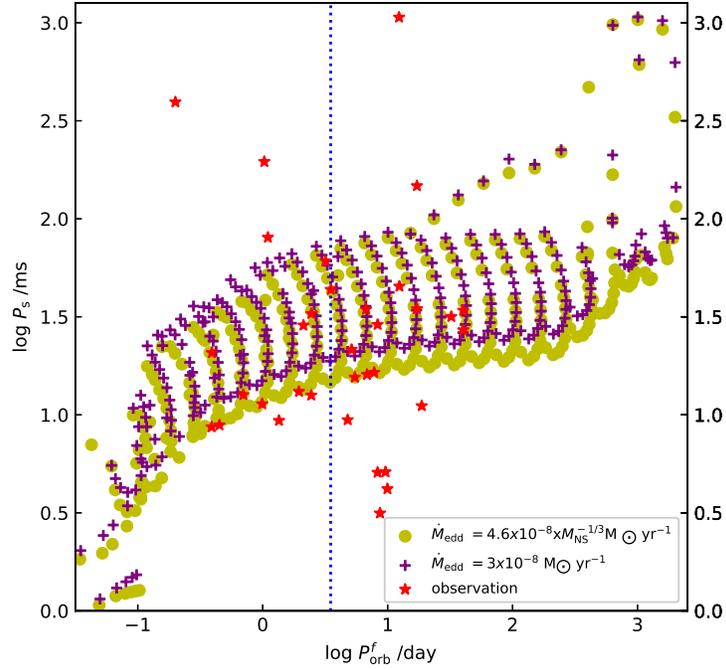,angle=0,width=10.5cm}
 \caption{Similar to Fig.\ 2,  but different descriptions of $\dot M_{\rm Edd}$ are adopted. The red stars stand for observed sources, purple crosses and yellow circles represent $\dot M_{\rm Edd}$ according to case 1 and $\dot M_{\rm Edd}$ according to case 2, respectively.}
  \end{center}
\end{figure}

We study the influence of different prescriptions of $\dot M_{\rm Edd}$ with case 1 and case 2 (see Fig.\ 3). The initial mass of NSs is set to be $1.30\,\rm M_{\odot}$ and Eq.\,\rm (4) is used to calculate $P_{\rm s}$. As shown in this figure, IMBPs in case 1 have shorter spin periods than those in case 2 under the same initial conditions. This is because that although they have almost the same mass transfer timescale, the $\dot M_{\rm Edd}$ in case 2 is larger than that in case 1, resulting in a larger mass accumulation rate of NSs in case 2. For example, for a system with $M^{\rm i}_{\rm He}$ of $1.0\,\rm M_{\odot}$ and $P^{\rm i}_{\rm orb}$ of 10\,\rm d, the mass transfer timescale is about $1.3\times 10^{\rm 5}\,\rm yr$ for both two cases. While the mass accretion rate of case 2 is about $4.2\times 10^{\rm -8}\,\rm M_{\odot}$/yr and that in case 1 is  $3.0\times 10^{\rm -8}\,\rm M_{\odot}$/yr. So the NS in case 1 can only accrete a mass of about 0.00393$\,\rm M_{\odot}$ then its spin period is about 21.66 ms, but the NS in case 2 can accrete a mass 0.00548$\,\rm M_{\odot}$ and its spin period is around 16.88 ms. We also note that case 2 can cover more observed IMBPs with $P_{\rm orb}$ less than 3\,\rm d on the parameter space of IMBPs.

\begin{figure}
\begin{center}
\epsfig{file=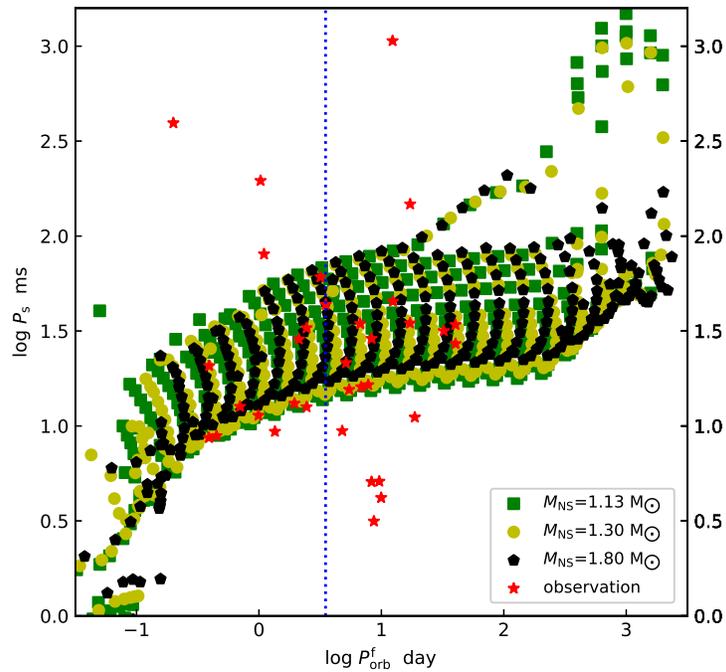,angle=0,width=10.5cm}
 \caption{The parameter space of ($\log P^{\rm f}_{\rm orb}$–$\log P_{\rm s}$) for different NS masses. The green squares, yellow circles and black pentagons represent the case of 1.13 $\rm M_{\odot}$, 1.30 $\rm M_{\odot}$, 1.80 $\rm M_{\odot}$ NS, respectively.}
  \end{center}
\end{figure}

We investigate the influence of different NS masses (see Fig.\ 4). The NS masses are set to 1.13$\,\rm M_{\odot}$, 1.30$\,\rm M_{\odot}$ and 1.80$\,\rm M_{\odot}$. In this figure, $\dot M_{\rm Edd}$ and $P _{\rm s}$ are calculated for case 2 and Eq.\,\rm (4), respectively. The boundaries are determined as follows. The He stars beyond the upper boundaries will evolve to NSs or black holes rather than WDs. The lower boundaries represent the minimum mass of He stars which can fill their Roche lobe and recycle the NSs at same $P^{\rm i}_{\rm orb}$.  The left boundaries are determined by the condition that the He stars fill their Roche lobe when they are at the He zero-age main sequence stage. The systems beyond the right boundaries will not undergo mass transfer because the He stars can’t fill their Roche-lobe. From this figure, we can see that, the final spin periods of NSs slightly decrease with the initial NS masses. This is caused by that $\dot M_{\rm Edd}$ increases with decrease of NS masses (see Eq.\,\rm (2)). This figure also shows that 11 of the 15 observed IMBPs with $P_{\rm orb}$ less than 3\,\rm d are located in the predicted parameter space of IMBPs, which, together with figures 2 and 3, indicates that the NS+He star channel is an important path for the formation of IMBPs with short orbital periods. In addition, nearly half of the observed IMBPs with long orbital periods are also covered by the predicted parameter space, which implies that the NS+He star scenario is a possible scenario for the formation of IMBPs with long orbital periods. 

\subsection{PSR J0621+1002}
\begin{figure*}
\epsfig{file=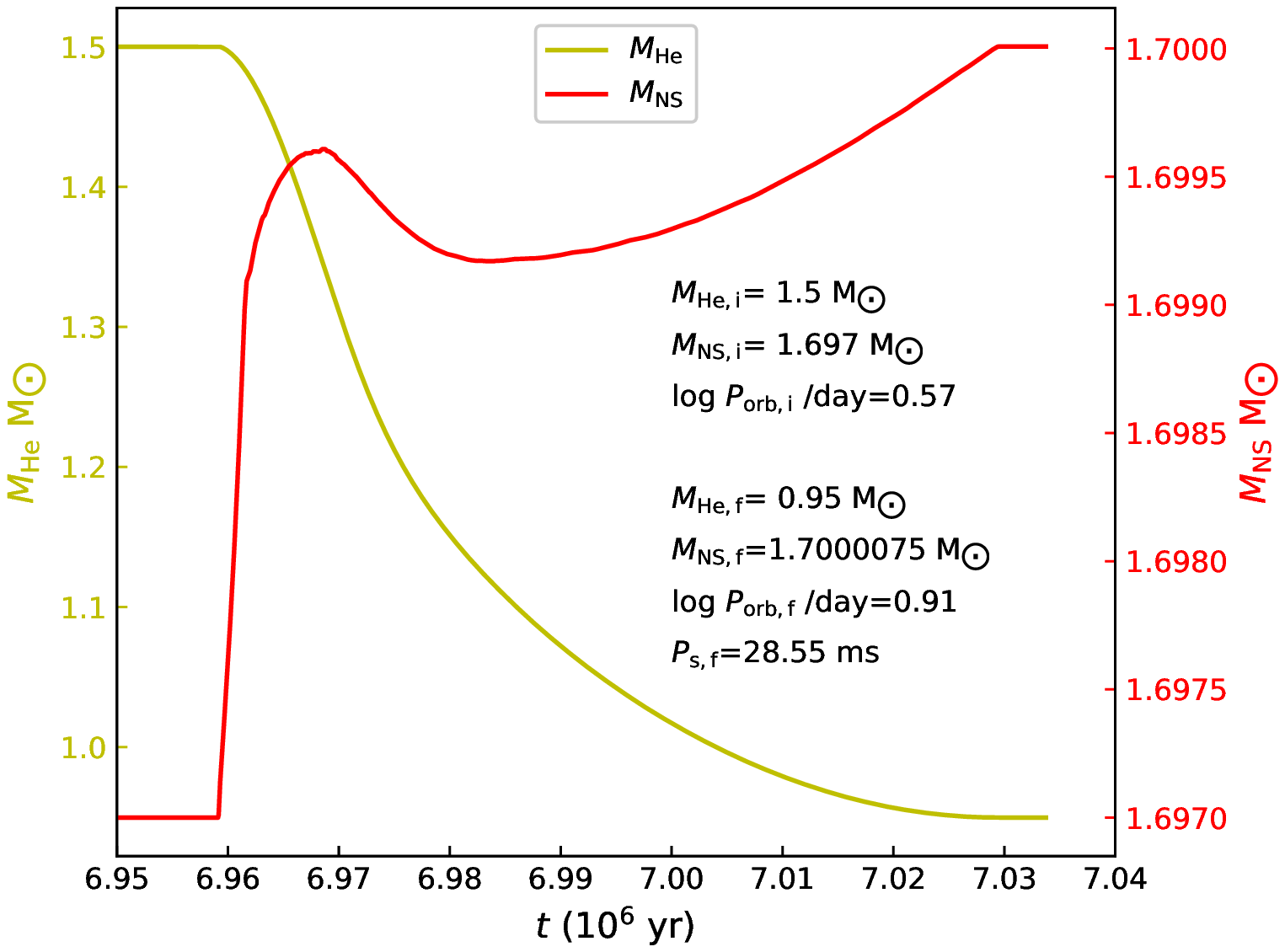,angle=0,width=10.5cm}\ \
\epsfig{file=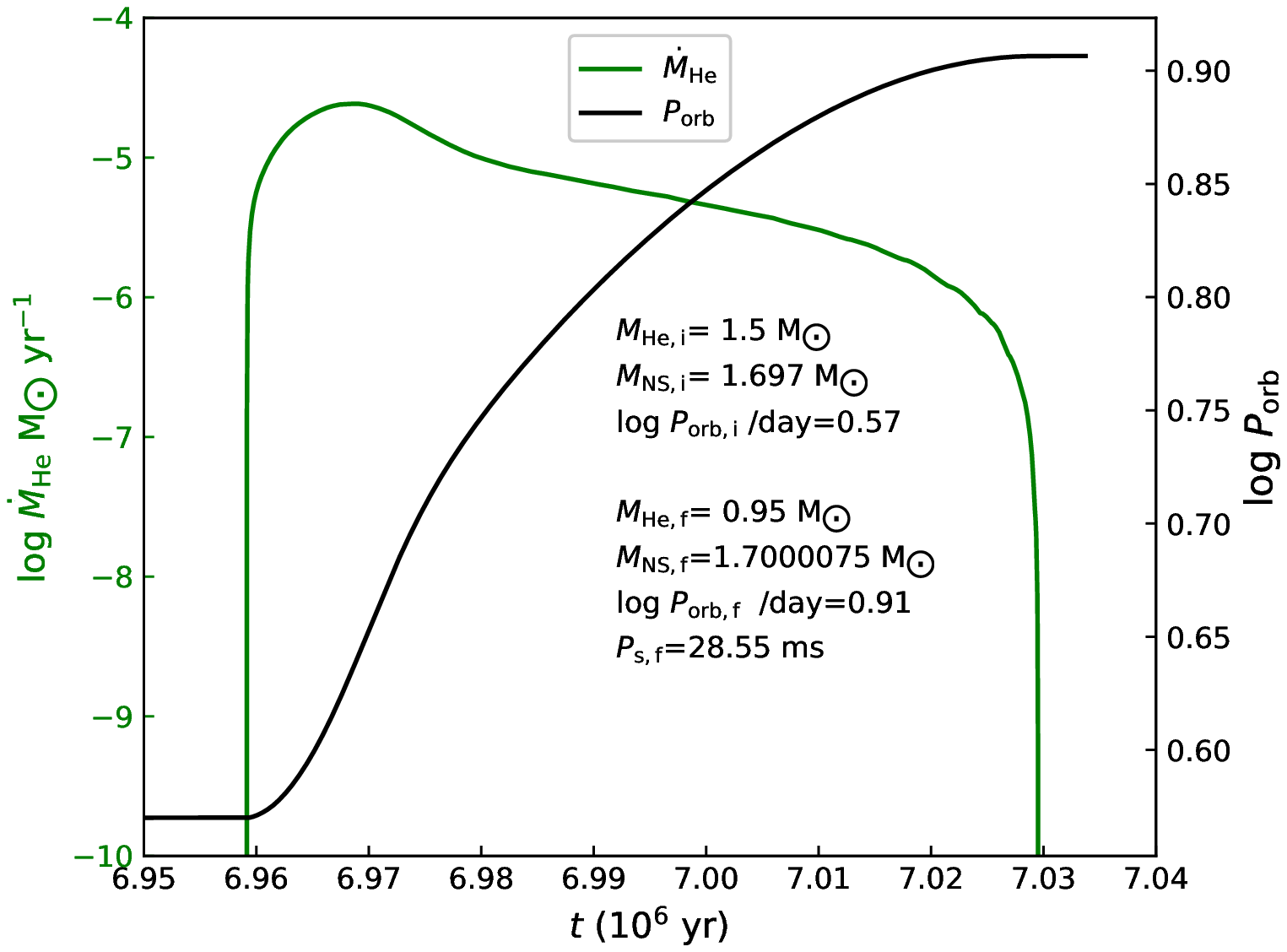,angle=0,width=10.5cm}
  \caption{A possible evolutionary path for the formation of PSR J0621+1002. In upper panel, the red and yellow curve represent the evolution of $M_{\rm NS}$ and $M_{\rm He}$; in lower panel, the green and black curve display the evolution of $\dot M_{\rm He}$ and $P_{\rm orb}$.}
\end{figure*}

The system PSR J0621+1002 is one of the well-observed IMBPs whose pulsar mass has been precisely measured. The system PSR J0621+1002 was found by Camilo et al. (1996) and it has a $1.70^{\rm +0.32}_{\rm -0.29}$$\,\rm M_{\odot}$ NS with spin period of 28.85\,\rm ms and a $0.97^{\rm +0.27}_{\rm -0.15}$$\,\rm M_{\odot}$ CO WD with orbital period of 8.32\,\rm d (Splaver et al. 2002). In this work, we reproduce the observed properties of this binary pulsar including the mass of the NS and companion star, the spin period of the NS and the orbital period. The detailed evolutionary path is shown in Fig.\ 5. The initial binary system consists of a 1.697$\,\rm M_{\odot}$ NS and a 1.5$\,\rm M_{\odot}$ He star with initial orbital period of about 3.72\,\rm d. The $\dot M_{\rm Edd}$ is adopted as case 2. The formed IMBP by this initial binary system includes a 1.7000075$\,\rm M_{\odot}$ NS with spin period of 28.55\,\rm ms and a 0.95$\,\rm M_{\odot}$ CO WD with orbital period of 8.13\,\rm d, which is in agreement with observed properties of PSR J0621+1002. Table 2 summarizes the observation and theoretical results of PSR J0621+1002.

\section{Discussions} \label{4. Discussion}

We found that different methods for calculating $P_{\rm s}$ have great influence on the final spin periods of NSs, depending on which accretion radius we use when calculating $P_{\rm s}$. For example, as shown in Sect. 3.1, after accreting a mass about 0.0051$\,\rm M_{\odot}$, the spin period of the NS is 17.82\,\rm ms and 47.19\,\rm ms when we use magnetospheric boundary and NS radius, respectively. For further discussion, we first introduce the magnetic radius ($\rm r_{m}$) and co-rotation radius ($\rm r_{co}$). At $\rm r_{\rm m}$, the magnetic and the infalling material stresses are equal, and the $\rm r_{co}$ is the position where the stellar and Keplerian angular velocities are equal and  the plasma is forced to corotate with the star by magnetic field inside $\rm r_{co}$ (e.g. Tauris, Langer \& Kramer 2012). According to Tauris, Langer \& Kramer (2012), in the equilibrium spin phase, $\rm r_{mag}$ is  approximately equal to $\rm r_{co}$. So the outermost boundary where stellar and accreted material have same angular velocity is at magnetospheric radius $\rm r_{m}$ when the pulsar achieves its equilibrium spin period ($P_{\rm eq}$). Therefore, we argue that it is also reasonable to use $\rm r_{\rm m}$ when we try to derive an analytical expression for the equilibrium spin period. Even so, however, we still argue that there is great uncertainty on how much matter needs to be accreted to spin up a NS to its current period. For example, Zhang et al. (2011) also proposed an empirical relation between the accreted mass ($\Delta M_{\rm NS}$) and its spin period as $\Delta M_{\rm NS}$ $\propto$ $P_{\rm s}$$^{\rm -2/3}$, which is different from the two methods mentioned above. Therefore, this should be further researched. 

In our calculations, we set the accretion efficiency $e_{\rm acc}\cdot k_{\rm def}=1$, which means that 100\% of the transferred matter would be accumulated on the surface of the NS if  $\dot{M}_{\rm 2}$ $\le$ $\dot{M}_{\rm Edd}$. However, it does not always work for all cases. For example, recent observations indicate that there exist inefficient accretion in LMXBs, even when $\dot{M}_{\rm 2} \leq \dot{M}_{\rm Edd}$ (e.g. Tauris et al. 1999; Jacoby et al. 2005; Antoniadis et al. 2012;). Some possible mechanisms (including propeller effects, accretion disc instabilities, and direct irradiation of the donor’s atmosphere from the pulsar) were supposed to account for inefficient accretion (e.g. Illarionov \& Sunyaev 1975; van Paradijs 1996; Dubus et al. 1999). On the other hand, the accretion rate also may be larger than the Eddington accretion rate. For example, it has been suggested that a NS may accrete matter at a rate of 2$\sim$3 $\dot{M}_{\rm Edd}$ in Case BB RLOF (Tauris, Langer \& Kramer 2012; Lazarus et al. 2014 ). So if considering more detailed real physics, the accretion efficiency will change and our results may be different as well.

 Another issue is how those massive NSs were born, or whether such massive NSs exist.	 Observationally, PSR J0348+0432 (Antoniadis et al. 2013), PSR J1614-2230 (Hessels et al. 2005; Demorest et al. 2010) both have an about $2\,\rm M_{\odot}$ NS, which indicates that the birth mass of NSs may be higher than $1.6\,\rm M_{\odot}$ (see Tauris et al. 2011; Antoniadis et al. 2013; Lin et al. 2011; Fortin et al. 2016). Recently, Cromartie et al. (2019) found that the system PSR J0740+6620 has an about 2.17$\,\rm M_{\odot}$ NS. Theoretically, by performing simulations of neutrino-powered explosions in spherical symmetry, Ugliano et al. (2012) found that  the birth masses of NSs vary from 1.2$-$2.0$\,\rm M_{\odot}$. Fortin et al. (2016) also investigated the possible scope of birth mass of NSs from SN explosion. Starting with a double MS binary system, they found that the birth mass of a NS could be as high as $1.9\,\rm M_{\odot}$, which subsequent evolution can be consistent with the observation of the system PSR J1614-2230. They concluded that the masses of progenitor NSs of recycled pulsars are in a broad interval varying from 1.0 $\rm M_{\odot}$ to 1.9$\,\rm M_{\odot}$. Both of these results indicate that the birth masses of NSs are in a broad region and our chosen NS masses are within this scope. Statistically, there may be three peaks for the NS masses distribution (1.25$\,\rm M_{\odot}$, 1.35$\,\rm M_{\odot}$,1.73$\,\rm M_{\odot}$; e.g. Schwab et al. 2010; Valentim et al. 2011). The NSs in the first peak originate from electron-capture supernovae (SNe) (e.g. Schwab et al. 2010). The NSs in latter two peaks may be formed from iron core collapse SNe of 10$-$20$\,\rm M_{\odot}$ stars and 20$-$25$\,\rm M_{\odot}$ stars, respectively (e.g. Tauris et al. 2011). Recent works show that there is quite a lot of randomness in the precise outcome of NS masses in SN collapse, as a function of initial main-sequence stellar mass (e.g. Sukhbold et al. 2016; Sukhbold, Woosley and Heger 2018).

\section{summary} \label{5.summary}

In this work, we investigate the NS+He star scenario for the formation of IMBPs, especially those IMBPs with $P_{\rm orb}$ less than 3\,\rm d. Our results are summarized as follows: 

(1) We have evolved a large number of NS+He star binaries and thus obtained a broad parameter space of IMBPs. The initial masses of NSs are set to be the range of about $1.10-1.80\,\rm M_{\odot}$; The initial masses of He stars are in the range of $0.67-2.20\,\rm M_{\odot}$ ; $\log P^{\rm i}_{\rm orb}$ /d changes from $-$1.3 to 3.3. 

(2) When calculating the spin period of a NS, it can be spun up to faster spin when using as accretion radius $\rm r_{\rm m}$ than with R.

(3) As we can see in Fig.\ 4, a less massive NS tends to be spun up to a smaller spin period value under the same initial conditions. On the other hand, the lower boundary of NS+He systems that can form IMBPs moves up when the initial NS mass increases.

(4) Our results can cover 11 of the 15 observed IMBPs with $P_{\rm orb}$$<$3\,\rm d in the $\log P^{\rm f}_{\rm orb}$$-$$\log P_{\rm s}$ plane. In addition, the observed properties of nearly half of the IMBPs with $P_{\rm orb}$$>$ 3\,\rm d can be covered by our results. Thus, we suggest that the NS+He star scenario is not only an important path to form the IMBPs with short orbital periods but also an alternative scenario to produce the IMBPs with long orbital periods.

(5) The observed properties of the system PSR J0621+1002 can be also quite well reproduced and we show a possible evolutionary history of PSR J0621+1002.

\section*{Acknowledgments}

We acknowledge useful comments and suggestions from the anonymous referee.This study is supported by the National Natural Science Foundation of China (Nos 11873085, 11673059 and 11521303), the Chinese Academy of Sciences (Nos QYZDB-SSW-SYS001 and KJZD-EW-M06-01 and Y9XB016001), and the Yunnan Province (Nos 2017HC018, 2018FB005).

\label{lastpage}
\end{document}